\documentclass[journal=jacsat,manuscript=article]{achemso}
\setkeys{acs}{articletitle = true}
\usepackage[version=3]{mhchem} % Formula subscripts using \ce{}
\usepackage[T1]{fontenc}       % Use modern font encodings
\usepackage{amssymb}
\author{Tokarchuk M.}
\email{mtok@icmp.lviv.ua}
\affiliation{Institute for Condensed Matter Physics of the National Academy of Sciences of Ukraine, \\ 1 Svientsitskii str., 79011, Lviv, Ukraine}
\altaffiliation{Lviv Polytechnic National University,  12 S. Bandera str., 79013, Lviv, Ukraine}

\title[Kinetic description of ion transport in the system ``ionic solution -- porous environment'']
  {Kinetic description of ion transport in the system ``ionic solution -- porous environment''}

\abbreviations{IR,NMR,UV}

\begin{document}

%%%%%%%%%%%%%%%%%%%%%%%%%%%%%%%%%%%%%%%%%%%%%%%%%%%%%%%%%%%%%%%%%%%%%%
%%% The "tocentry" environment can be used to create an entry for the
%%% graphical table of contents. It is given here as some journals
%%% require that it is printed as part of the abstract page. It will
%%% be automatically moved as appropriate.
%%%%%%%%%%%%%%%%%%%%%%%%%%%%%%%%%%%%%%%%%%%%%%%%%%%%%%%%%%%%%%%%%%%%%%
%\begin{tocentry}
%
%Some journals require a graphical entry for the Table of Contents.
%This should be laid out ``print ready'' so that the sizing of the
%Inside the \texttt{tocentry} environment, the font used is Helvetica
%8\,pt, as required by \emph{Journal of the American Chemical
%The surrounding frame is 9\,cm by 3.5\,cm, which is the maximum
%permitted for  \emph{Journal of the American Chemical Society}
%graphical table of content entries. The box will not resize if the
%content is too big: instead it will overflow the edge of the box.
%
%This box and the associated title will always be printed on a
%separate page at the end of the document.
%
%\end{tocentry}

\begin{abstract}
  A kinetic approach based on a modified chain of BBGKI equations for nonequilibrium particle distribution
           functions was used to describe the ion transfer processes in the ionic solution -- porous medium system.
           A generalized kinetic equation of the revised  Enskog--Vlasov--Landau theory for the nonequilibrium ion
           distribution function in the model of charged solid spheres is obtained,
           taking into account attractive short-range interactions for the ionic solution -- porous medium system.
\end{abstract}

\section{Introduction}
 Studies of particle transfer processes (atoms, molecules, ions) in porous media
 (soils, nano heterogeneous composite materials -- electrodes, polymer, and biological membranes, etc.) are relevant in terms of modern technologies in the creation of nano batteries, supercapacitors, membrane structures for separation purification of aqueous solutions, etc.
 The importance and justification of such studies are noted in many works,
 including~\cite{Sahimi,Hatano,Gelb,lit1,lit3,Berkowitz,Berkowitz1,Smith,Neuman,Roten,Yang,Maf,Bijeljic,6,lit5,Tyukhova,Comolli,Waisbord,Zhao}. The study of the mechanisms of anomalous dispersion of rheological fluid flows in inhomogeneous porous media~\cite{Yang,Tyukhova,Comolli,Waisbord} is relevant from the point of view of practical applications.
 Another aspect of research is related to the ionic conductivity of ionic solutions in porous and layered structures,
 which is important in connection with the anomalous behavior of ion diffusion and polarization effects~\cite{Bisquert2001112,Hatano,Sibatov1,Sibatov2,Nigmatullin,6}.
 In particular,
 theoretical studies of electrodiffusion processes of ion transfer in systems are relevant ``electrolyte -- electrode''~\cite{lit1,lit3,6,lita5,lita5b,lit6,lit7,uch1,kost2,lita3,lita6} and associated with the need to describe nonequilibrium processes of intercalation--deintercalation of ions,
 and with the need for a theory suitable for practical application for forecasting and management these processes.
 Problems in the description of electrode processes are connected,
 first of all,
 as with the superficial phenomena at the interface ``electrolyte -- electrode'',
 where complex processes of adsorption,
 desorption,
 diffusion,
 and in the middle of the electrodes (porous, layered in structure),
 where complex processes of association,
 dissociation between ions,
 their anomalous (sub or super) diffusion due to complex interaction with the structure of the electrode,
 which in turn are associated with problems of charge accumulation on electrodes in batteries.
 Therefore,
 it is very important to take into account,
 to some extent,
 the change in microstructure electrode material,
 in particular,
 due to its polarization properties and porosity.
 In the vast majority of studies to describe electrodiffusion processes of ion transfer in systems ``electrolyte -- electrode'' equations of nonequilibrium thermodynamics~\cite{lit6} with constant diffusion coefficients are used.

 It is interesting to study the processes of self-diffusion of ions in charged nanoporous media (particles which are frozen) by computer simulation methods~\cite{Jardat1,Jardat2}.
 At the same time,
 an important feature of these systems is their significant spatial heterogeneity,
 when the diffusion coefficients are functions of spatial coordinates and time,
 that is,
 temporal correlation functions ``stream -- stream''
 $\langle\hat{\vec{j}}(\vec{r}_{l};t)\hat{\vec{j}}(\vec{r}_{l'};t')\rangle$
 in each of the phases and between phases.
 It should be noted that some way of calculating the diffusion coefficients of ions depending on the coordinates for the systems ``electrolyte solution -- membrane'',
 ``electrolyte solution -- vitreous fuel-containing materials'',
 ``electrolyte solution -- soil'' was proposed in~\cite{Omelyan,Yukhnovskii,Yukhnovskii1}.
 In~\cite{6,kost2},
 a statistical theory was proposed to describe electrodiffusion ion transfer processes in the system ``electrolyte -- electrode '' taking into account the spatial inhomogeneities and memory effects using the nonequilibrium method statistical operator (NSO).
 In~\cite{6,lita3,lita5,lita6}, experimental and theoretical study of subdiffusion impedance for multilayer system GaSe with encapsulated $\beta$-cyclodextrin having a porous fractal structure was carried out.
 From the point of view of theoretical research,
 the generalized equations of electrodiffusion of Cattaneo type in fractional derivatives \cite{lita6} were applied.
 However,
 in most of the mentioned works,
 except for~\cite{Omelyan,Yukhnovskii,Yukhnovskii1},
 in the corresponding transfer equations,
 the transfer coefficients: diffusion, viscosity, and thermal conductivity,
 which determine the main mechanisms of transfer processes,
 are constants at the corresponding temperatures.
 Obviously,
 to elucidate such mechanisms,
 it is very important to apply such theoretical approaches that link the transfer coefficients with the characteristic potentials of the interaction of particles and their distribution functions.

 In this paper,
 we will apply the kinetic approach to the description of ion transfer processes in the system ionic solution -- porous medium.
 The second section analyzes the modified chain of BBGKI equations for nonequilibrium ion distribution functions in the ionic solution -- porous medium system,
 depending on whether the two subsystems interacted at the initial time.
 The third section considers the chain of BBGKI equations for nonequilibrium ion distribution functions in the pairwise collision approximation for the case when the ionic solution and porous matrix subsystems are interacting.
 For this case,
 a generalized kinetic equation of the revised Enskog--Vlasov--Landau theory for the nonequilibrium ion distribution function in the model of charged solid spheres will be obtained,
 taking into account attractive short-range interactions for the ionic solution-porous medium system.
 To calculate the paired quasi-equilibrium coordinate distribution functions for ions and particles of the porous matrix,
 inhomogeneous,
 time-dependent Ornstein--Zernike equations will be proposed for the corresponding complete correlation functions.

  \section{The chain of BBGKI equations for the ionic solution system in a porous medium}

 We consider a system of ionic solution that interacts with a porous medium by diffusing at its place.
 Positively and negatively charged ions of the solution can penetrate into the structure of the porous medium (matrix) and move in it,
 interacting with its particles.
 We assume the whole volume of the system to be equal to $V=V_{l}+V_{s}$,
 where $V_{l}$ is the volume occupied by the ionic solution,
 and $V_{s}$ is the true volume porous matrix.
 Entering the volume $V_{por}$ of the porous space of the matrix,
 we can determine its porosity:
 $\psi = 1-\frac{V_{por}}{V}$.

 The ionic solution will be considered with specific dielectric properties without explicit consideration molecular subsystem,
 and the porous matrix formed by moving particles (atoms, molecules) varieties $\xi$ subsystem,
 kinetic energy,
 which is much less than its potential energy.
 We believe that the particles of the porous matrix perform oscillating motions,
 and its dynamics can be described by interacting phonons,
 for example,
 in the case of porous silicon,
 and other porous materials~\cite{Alvarez}.
 When interacting with an ionic solution,
 both the structure of the porous matrix and the dynamic and structural properties of the ionic solution itself can change.
 The Hamiltonian of such a system is an ionic solution -- a porous matrix can be represented as:\vspace{-2mm}
 \begin{align*}%\label{e.01}
  H&=\sum_{\alpha,j=1}^{N_{\alpha}}\frac{\big(\vec{p}_{j}-\frac{Z_{\alpha}e}{c}\vec{A}(\vec{r}_{j};t)\big)^{2}}{2m_{\alpha}} +\sum_{\alpha,\gamma}\sum_{j,l}^{N_{\alpha},N_{\gamma}}\Phi_{\alpha\gamma}(\vec{r}_{j},\vec{r}_{l})\nonumber\\
   &\quad+\sum_{\alpha,\xi}\sum_{j,l}^{N_{\alpha},N_{\xi}}
     \Phi_{\alpha \xi}(\vec{r}_{j},\vec{R}_{l})+\sum_{\alpha}\sum_{j}^{N_{\alpha}}Z_{\alpha}e\varphi(\vec{r}_{j};t)+H_{s},
 \end{align*}
 where the indices $j$, $l$ number the ions of the solution of varieties $\alpha$, $\gamma$,
 with masses $m_{\alpha}$, $m_{\gamma}$ and vector--momentum $\vec{p}_{j}$, $\vec{p}_{l}$, $\vec{r}_{j}$, $\vec{R}_{l}$ are the coordinate vectors of ions and particles,
 respectively porous matrix,
 $N_{\alpha}$ is the total number of ions of the variety $\alpha$;
 $\Phi_{\alpha\gamma}(\vec{r}_{j},\vec{r}_{l})=\Phi^{sh}_{\alpha\gamma}(\vec{r}_{j},\vec{r}_{l})+\Phi^{l}_{\alpha\gamma}(\vec{r}_{j},\vec{r}_{l})$ is an even potential for interaction between ions of the grade $\alpha$, $\gamma$,
 $\Phi_{\alpha \xi}(\vec{r}_{j},\vec{R}_{l})=\Phi^{sh}_{\alpha \xi}(\vec{r}_{j},\vec{R}_{l})+\Phi^{l}_{\alpha \xi}(\vec{r}_{j},\vec{R}_{l})$ is the pairwise potential of ion interaction
 with porous matrix particles that have short-range and long-range (attractive) contributions.

 Moreover,
 short-term contributions can have repulsive and attractive components $\Phi^{sh}_{\alpha\gamma}(\vec{r}_{j},\vec{r}_{l})=\Phi^{sh-rep}_{\alpha\gamma}(\vec{r}_{j},\vec{r}_{l})+\Phi^{sh-att}_{\alpha\gamma}(\vec{r}_{j},\vec{r}_{l})$, i.e.~repulsive and attractive forces can act between ions at small distances.
 In particular, attractive short-range interactions can describe the processes of association between ions. $H_{s}$ is a Hamiltonian of a porous matrix, the structure of which will not be specified at this stage of modeling. $\vec{A}(\vec{r}_{j};t)$,
 $\varphi(\vec{r}_{j};t)$ are the total vector and scalar potentials of the electromagnetic field generated by valence ions $Z_{\alpha}$ and external field,
 $e$ is the electron charge and $c$ is the speed of light.
 In what follows,
 we will not consider vortex electromagnetic processes of order $\frac{e}{c}$,
 but only potential contributions from the scalar potential $\varphi(\vec{r}_{j};t)$.
 In addition,
 at this stage of research we do not take into account the possible influence of the electromagnetic field on the particles of the porous matrix,
 which may be associated with the processes of polarization,
 changes in its dielectric properties.
 Although in reality,
 in many cases,
 in particular for electrodes,
 biological membranes,
 such consideration is important.

 The nonequilibrium state of the ionic solution when interacting with the porous matrix is completely described by the Liouville equation for nonequilibrium distribution function of all particles
 $\rho(x_{1},\ldots,x_{N_{\alpha}}|\vec{R}_{1},\ldots,\vec{R}_{N_{\xi}};t)=\rho(x^{N_{\alpha}},R^{N_{\xi}};t)$
 (where the notation is entered: $x^{N_{\alpha}}=x_{1},\ldots,x_{N_{\alpha}}$,
 $x_{j}=\vec{p}_{j}\vec{r}_{j}$ is the coordinate and momentum of $j$-th ion of the solution, $R^{N_{\xi}}=\vec{R}_{1},\ldots,\vec{R}_{N_{\xi}}$,
 $\vec{R}_{l}$ is the coordinate of the $l$-th particle of the porous matrix)\vspace{-2mm}
 \begin{equation*}%\label{e.001}
  \frac{\partial}{\partial t}\rho(x^{N_{\alpha}},R^{N_{\xi}};t)+iL_{N}\rho(x^{N_{\alpha}},R^{N_{\xi}};t)=0
 \end{equation*}\\[-6mm]
 with the Liouville operator\vspace{-3mm}
 \begin{align*} %\label{e.02}
  iL_{N}&=\sum_{\alpha,j=1}^{N_{\alpha}}\frac{\vec{p}_{j}}{m_{\alpha}}\cdot\frac{\partial }{\partial \vec{r}_{j}}+\sum_{\xi,l=1}^{N_{\xi}}\frac{\vec{P}_{l}}{m_{\xi}}\cdot \frac{\partial }{\partial \vec{R}_{l}}
  -\sum_{\alpha,\gamma}\sum_{j,l}^{N_{\alpha}N_{\gamma}}\frac{\partial }{\partial \vec{r}_{j}}\Phi_{\alpha\gamma}(\vec{r}_{j},\vec{r}_{l})
  \left(\frac{\partial }{\partial \vec{p}_{j}}- \frac{\partial }{\partial \vec{p}_{l}}\right) \nonumber\\[-1mm]
  &\quad- \sum_{\alpha,\xi}\sum_{j,l}^{N_{\alpha},N_{\xi}}
   \frac{\partial }{\partial \vec{r}_{j}}
   \left(\Phi_{\alpha \xi}(\vec{r}_{j},\vec{R}_{l})+Z_{\alpha}e\varphi(\vec{r}_{j};t)\right)\cdot
  \frac{\partial }{\partial \vec{p}_{j}},
 \end{align*}\\[-5mm]
 at the same time in the Liouville operator we do not consider terms %\vspace{-3mm}
 \(
  -
  \sum_{\alpha,\xi}\sum_{j,l}^{N_{\alpha},N_{\xi}}\frac{\partial }{\partial \vec{R}_{l}}
  \Phi_{\alpha \xi}(\vec{r}_{j},\vec{R}_{l})\cdot \frac{\partial }{\partial \vec{P}_{l}},
 \) %\\[-4mm]
 since we believe that the microscopic forces acting on the atoms of the porous matrix on the ion of the solution do not change the momentum of the particles of the matrix.

 The nonequilibrium state of the system ionic solution -- porous medium will be described using a modified chain of BBGKI equations~\cite{l1,tokom,zmot,tokom01,tokom02} for partial nonequilibrium distribution functions of ions and particles of a porous matrix.
 To do this,
 we use the approach proposed in the works~\cite{l1,tokom,zmot,tokom01,tokom02},
 where the modified chain of equations BBGKI is built,
 taking into account the concept of a consistent description of kinetics and hydrodynamics of nonequilibrium processes of the system of interacting particles in the method of nonequilibrium statistical operator Zubarev,
 based on the Liouville equation with the source:\vspace{-2mm}
 \begin{equation}       \label{e.0010}
  \frac{\partial }{\partial t}\rho\left(x^{N_{\alpha}},R^{N_{\xi}};t\right)
  +iL_{N}\rho\left(x^{N_{\alpha}},R^{N_{\xi}};t\right)
  =
  -\varepsilon\left(\rho\left(x^{N_{\alpha}},R^{N_{\xi}};t\right)-\varrho_{rel}\left(x^{N_{\alpha}},R^{N_{\xi}};t\right)\right),
 \end{equation}\\[-6mm]
 which selects the delayed
 ($\varepsilon\rightarrow +0$,
 after thermodynamic limit)
 solutions of the Liouville equation under given initial conditions.

 In this case, the initial condition for solving the Liouville equation (Cauchy problem) can be considered two options: \textbf{first}\vspace{-2mm}
 \[
  \rho\left(x^{N_{\alpha}},R^{N_{\xi}};t\right)\big|_{t=t_{0}}
  =
  \varrho_{\rm rel}\left(x^{N_{\alpha}},R^{N_{\xi}};t_{0}\right)
  =
  \rho_{\rm rel}^{\rm liq}\left(x^{N_{\alpha}};t_{0}\right)\rho_{s}\left(R^{N_{\xi}};t_{0}\right),
 \]
 this means that at the initial moment of time $t_{0}$ the ionic solution and the porous matrix are considered independent; \textbf{other}
 \[
  \rho\left(x^{N_{\alpha}},R^{N_{\xi}};t\right)\big|_{t=t_{0}}
  =
  \rho_{\rm rel}\left(x^{N_{\alpha}},R^{N_{\xi}};t_{0}\right),
 \]
 when the ionic solution and the porous matrix are considered at the initial moment of time by interacting subsystems.

 In the \textbf{first} case, $\rho_{\rm rel}^{\rm liq}(x^{N_{\alpha}};t)$ is the relevant ion distribution function obtained by~\cite{l1,tokom,zmot,tokom01,tokom02} from the condition of the maximum of the Gibbs entropy functional while preserving the normalization conditions for the distribution and the given parameters of the abbreviated description of the nonequilibrium state of the ionic solution:
 $\langle\hat{n}_{\alpha}(x)\rangle^{t}=f_{\alpha}(x;t)$ is the nonequilibrium one-particle ion distribution function of the variety $\alpha$ and $\langle\hat{\varepsilon}_{\rm int}(\vec{r})\rangle^{t}$ is the nonequilibrium. The average interaction energy of the ions of the solution has the following structure:\vspace{-2mm}
 \begin{equation*}%\label{e.002}
  \rho_{\rm rel}^{liq}(x^{N_{\alpha}};t)
  =
  \exp\left\{-\Phi_{\rm liq}(t)
  -\int d\vec{r}\,\beta_{\rm liq}(\vec{r},t)\hat{\varepsilon}_{\rm int}(\vec{r})
  -\sum_{\alpha}\int dx\,a_{\alpha}(x,t)\hat{n}_{\alpha}(x)\right\},
 \end{equation*}\\[-4mm]
 where $\Phi_{\rm liq}(t)$ is the Masier--Planck functional\vspace{-3mm}
 \begin{equation*}%\label{e.003}
  \Phi_{\rm liq}(t)
  =
  \int d\Gamma_{N_{\rm liq}}(x)\,
  \exp\left\{-\int d\vec{r}\,\beta_{\rm liq}(\vec{r},t)\hat{\varepsilon}_{\rm int}(\vec{r})
  -\sum_{\alpha}\int dx \, a_{\alpha}(x,t)\hat{n}_{\alpha}(x)\right\} ,
 \end{equation*}\\[-4mm]
 in which
 \begin{equation*}      % \label{e.004}
  \hat{n}_{\alpha}(x)
  =
  \sum_{j=1}^{N_{\alpha}}\delta (x-x_{j})
  =
  \sum_{j=1}^{N_{\alpha}}\delta(\vec{r}-\vec{r}_{j})\delta(\vec{p}-\vec{p}_{j})
 \end{equation*}\\[-3mm]
 is the microscopic phase density of the number of ions of the variety $\alpha$ and\vspace{-2mm}
 \begin{equation*}       %\label{e.005}
  \hat{\varepsilon}_{\rm int}(\vec{r})
  =
  \frac{1}{2}\sum_{\alpha,\gamma}\sum_{j,l}^{N_{\alpha},N_{\gamma}}
  \Phi_{\alpha\gamma}(\vec{r}_{j},\vec{r}_{l})
  \delta (\vec{r}-\vec{r}_{j})
 \end{equation*}\\[-5mm]
 is the microscopic energy density of the interaction of the ions of the solution.
 The Lagrange parameters $\beta_{\rm liq}(\vec{r},t)$
 (inverse of the nonequilibrium temperature of the ionic solution),
 $a_{\alpha}(x,t)$ are determined from the conditions of self-consistent:\vspace{-2mm}
 \[
  \langle\hat{\varepsilon}_{\rm int}(\vec{r})\rangle^{t}
  =
  \langle\hat{\varepsilon}_{\rm int}(\vec{r})\rangle^{t}_{\rm rel}, \quad
  \langle\hat{n}_{\alpha}(x)\rangle^{t}
  =
  \langle\hat{n}_{\alpha}(x)\rangle^{t}_{\rm rel}.
 \]\\[-7mm]
 In this case,
 the ionic solution has a temperature $\beta_{\rm liq}(\vec{r},t)$,
 and the porous matrix $\beta_{s}$.

 In the \textbf{second} case, $\rho_{\rm rel}(x^{N_{\alpha}},R^{N_{\xi}};t)$ is the relevant function of the distribution of ions and particles of the porous matrix obtained respectively~\cite{l1,tokom,zmot,tokom01,tokom02} from the condition of the maximum of the Gibbs entropy functional with preserved normalization conditions for the distribution and set parameters of the abbreviated description of the nonequilibrium state of the ionic solution: $\langle\hat{n}_{\alpha}(x)\rangle^{t}=f_{\alpha}(x;t)$ is nonequilibrium one-particle ion distribution function of $\alpha$,
 $\langle \hat{E}_{int}(\vec{r})\rangle^{t}$ is nonequilibrium average ion interaction energy solution and particles of a porous matrix,
 $\langle\hat{n}_{\xi}(\vec{R})\rangle^{t}=n_{\xi}(\vec{R};t)$ is average the particle density of the porous matrix of the variety $\xi$ has the following structure:\vspace{-3mm}
 \begin{multline*}%\label{e.0021}
  \rho_{\rm rel}(x^{N_{\alpha}},R^{N_{\xi}};t)\\
  =
  \exp\left\{-\Phi(t)-\int \!\!d\vec{r}\,\beta(\vec{r},t)\big(\hat{E}_{int}(\vec{r})+H_{s}\big)
  -
  \sum_{\alpha}\int \!\!dx\, a_{\alpha}(x,t)\hat{n}_{\alpha}(x)
  -
  \sum_{\xi}\int \!\! d\vec{R}\,\mu_{\xi}(\vec{R})\hat{n}_{\xi}(\vec{R})\right\},
 \end{multline*}\\[-4mm]
 where  $\Phi(t)$ is a Masier--Planck functional\vspace{-3mm}
 \begin{multline*}%\label{e.0031}
  \Phi(t)=\int d\Gamma(x,R) \\
  \times\exp\left\{-\int\!\!d\vec{r}\,\beta(\vec{r},t)\big(\hat{E}_{int}(\vec{r})+H_{s}\big)
  -\sum_{\alpha}\int\!\!dx\, a_{\alpha}(x,t)\hat{n}_{\alpha}(x)
  -\sum_{\xi}\int\!\!d\vec{R}\,\mu_{\xi}(\vec{R})\hat{n}_{\xi}(\vec{R})\right\},
 \end{multline*}
 in which $\hat{n}_{\alpha}(x)$ is a microscopic phase density of the number of ions of the variety $\alpha$ and\vspace{-3mm}
 \begin{equation*}      % \label{e.0051}
  \hat{E}_{\rm int}(\vec{r})
  =
  \frac{1}{2}\sum_{\alpha,\gamma}\sum_{j,l}^{N_{\alpha},N_{\gamma}}\Phi_{\alpha\gamma}(\vec{r}_{j},\vec{r}_{l}) \delta (\vec{r}-\vec{r}_{j})
  +
  \sum_{\alpha,\xi}\sum_{j,l}^{N_{\alpha},N_{\xi}}\Phi_{\alpha \xi}(\vec{r}_{j},\vec{r}_{l}) \delta (\vec{r}-\vec{r}_{j})
 \end{equation*}\\[-5mm]
 is the microscopic energy density of the interaction of solution ions and particles of the porous matrix,
 which can perform oscillating motions.
 The Lagrange parameters $\beta(\vec{r},t)$
 (inverse of the nonequilibrium temperature of the ionic solution system -- porous matrix),
 $a_{\alpha}(x,t)$ are determined from the conditions of self-agreement:\vspace{-3mm}
 \[
  \big\langle(\hat{E}_{\rm int}(\vec{r})+H_{s})\big\rangle^{t}
  =
  \big\langle(\hat{E}_{\rm int}(\vec{r})+H_{s})\big\rangle^{t}_{\rm rel}, \quad
  \big\langle\hat{n}_{\alpha}(x)\big\rangle^{t}
  =
  \big\langle\hat{n}_{\alpha}(x)\big\rangle^{t}_{\rm rel}.
 \]\\[-7mm]
 In this case,
 the ionic solution and the porous matrix have a temperature of $\beta(\vec{r},t)$.

 \subsection{Kinetic equations with initial condition of independent subsystems: \\ ionic solution and porous matrix}

 Given the structure $\rho_{\rm rel}^{\rm liq}(x^{N_{\alpha}};t)$ and the approach~\cite{l1,tokom,zmot,tokom01,tokom02,kobr2,TokarchukN},
 integrating the Liouville equation with the source~(\ref{e.0010}) by the corresponding coordinates and momentum of the ions of the solution and the coordinates of the particles of the porous matrix,
 we obtain a chain of equations BBGKI with modified boundary conditions
 (taking into account spatiotemporal interparticle correlations)
 for the system ionic solution -- porous matrix:\vspace{-2mm}
 \begin{equation}\label{e.03}
  \left(\frac{\partial }{\partial t} +iL_{\alpha}(1)\right)f_{\alpha}(x_{1};t)
  + \sum_{\gamma}\int \!\!dx_{2}\,iL_{\alpha\gamma}(1,2) f_{\alpha\gamma}(x_{1},x_{2};t)
  + \int \!\!d\vec{R}_{s}\,iL_{\alpha s}(1,s) f_{\alpha s}(x_{1},\vec{R}_{s};t)=0,
 \end{equation}\\[-13mm]
 \begin{multline} \label{e.04}
  \left(\frac{\partial }{\partial t}
  +iL_{\alpha}(1)+iL_{\gamma}(2)+iL_{\alpha\gamma}(1,2)\right)
  f_{\alpha\gamma}(x_{1},x_{2};t) \\
  + \sum_{\nu}\int\!\! dx_{3} \big(iL_{\alpha\xi}(1,3) +iL_{\gamma\xi}(2,3)\big)
  f_{\alpha\gamma\nu}(x_{1},x_{2},x_{3};t)
  + \int\!\! d\vec{R}_{s}\big(iL_{\alpha s}(1,s) +iL_{\gamma s}(2,s)\big)
  f_{\alpha\gamma s}(x_{1},x_{2},\vec{R}_{s};t) \\
  =-\varepsilon \big(f_{\alpha\gamma}(x_{1},x_{2};t)
   -g_{\alpha\gamma}(\vec{r}_{1},\vec{r}_{2}|n,\beta;t)
   f_{\alpha}(x_{1};t)f_{\gamma}(x_{2};t)\big),
 \end{multline}\\[-16mm]
 \begin{multline} \label{e.04a}
  \left(\frac{\partial }{\partial t} +iL_{\alpha}(1)+iL_{\alpha s}(1,s)\right)
  f_{\alpha s}(x_{1},\vec{R}_{s};t) \\
  +\sum_{\gamma}\int dx_{3}\big(iL_{\alpha\gamma}(1,3)  +iL_{s\gamma}(s,3)\big)
   f_{\alpha s \gamma}(x_{1},\vec{R}_{s},x_{3};t)
  +\int d\vec{R}_{s'}iL_{\alpha s'}(1,s') f_{\alpha ss'}(x_{1},\vec{R}_{s},\vec{R}_{s'};t) \\
  =-\varepsilon \big( f_{\alpha s}(x_{1},\vec{R}_{s};t)- f_{\alpha}(x_{1};t)n(\vec{R}_{s};t)\big),
 \end{multline}\\[-8mm]
 where $\varepsilon \rightarrow +0$ after the thermodynamic limit,\vspace{-2mm}
 \begin{align} \label{e.05}
  iL_{\alpha}(j) &=\frac{\vec{p}_{j}}{m_{\alpha}}\cdot
  \frac{\partial }{\partial \vec{r}_{j}}-\frac{\partial }{\partial \vec{r}_{j}}
  Z_{\alpha}e\varphi(\vec{r}_{j};t)\cdot \frac{\partial }{\partial \vec{p}_{j}}, \\ \nonumber
  iL_{\alpha\gamma}(j,l) &= -\frac{\partial }{\partial  \vec{r}_{j}}
  \Phi_{\alpha\gamma}(\vec{r}_{j},\vec{r}_{l})\cdot
  \left(\frac{\partial }{\partial \vec{p}_{j}}- \frac{\partial }{\partial \vec{p}_{l}} \right)
 \end{align}\\[-5mm]
 there are one-part and two-part parts of the Liouville operator,\vspace{-2mm}
 \begin{equation*} %\label{e.06}
  iL_{\alpha s}(j,s)=-\frac{\partial }{\partial \vec{r}_{j}}
  \Phi_{\alpha s}(\vec{r}_{j},\vec{R}_{s})
  \cdot \frac{\partial }{\partial \vec{p}_{j}}
 \end{equation*}\\[-5mm]
 is a two-particle Liouville operator of particles of liquid subsystem and porous subsystem.
 $g_{\alpha\gamma}(\vec{r}_{1},\vec{r}_{2}|n,\beta;t)$ are quasi-equilibrium pair coordinate functions of ion distribution of varieties $\alpha,\gamma$ solution subsystems\vspace{-2mm}
 \begin{align*} %\label{e.0606}
  g_{\alpha\gamma}(\vec{r}_{1},\vec{r}_{2}|n,\beta;t)
  &=
  \frac{1}{n_{\alpha}(\vec{r}_{1};t)n_{\gamma}(\vec{r}_{2};t)}
  \int d\Gamma_{N_{\rm liq}}(x)\,\hat{n}_{\alpha}(\vec{r}_{1})
  \hat{n}_{\gamma}(\vec{r}_{2})\rho_{\rm rel}^{\rm liq}(x^{N_{\alpha}};t),\\ %\label{e.0607}
  n_{\alpha}(\vec{r};t)&=\int d\vec{p}\,f_{\alpha}(x;t)
 \end{align*}
 is the nonequilibrium average value of the density of the number of ions of the variety $\alpha$,
 $n(\vec{r}_{s};t)$ is a nonequilibrium average density of the number of particles of the porous subsystem, $f_{\alpha}(x_{1};t)$,
 $f_{\alpha\gamma}(x_{1},x_{2};t)$,
 $f_{\alpha\gamma\nu}(x_{1},x_{2},x_{3};t)$
 are one-, two- and three-ionic nonequilibrium distribution functions,
 $f_{\alpha\gamma s}(x_{1},x_{2},\vec{R}_{s};t)$,
 $f_{\alpha s}(x_{1},\vec{R}_{s};t)$,
 $f_{\alpha ss'}(x_{1},\vec{R}_{s},\vec{R}_{s'};t)$ are nonequilibrium distribution functions ions and particles of the porous medium.

 The three-particle functions satisfy the following equations of the chain of equations BBGKI,
 which include four-particle nonequilibrium distribution functions.
 It is important to note that one- and two-particle (ionic) nonequilibrium distribution functions determine the behavior of hydrodynamic variables: average non-equilibrium values densities of the number of ions $n_{\alpha}(\vec{r};t)$,
 their momentum $\vec{p}_{\alpha}(\vec{r};t)$,
 kinetic energy $\varepsilon^{\rm kin}_{\alpha}(\vec{r};t)$,
 as well as the potential energy $\varepsilon^{int}_{\alpha}(\vec{r};t)$:\vspace{-2mm}
 \begin{align*} %\label{e.070}
  \vec{p}_{\alpha}(\vec{r};t)&=\int\!\!d\vec{p}\,f_{\alpha}(\vec{r},\vec{p};t)\vec{p},\\ \label{e.071}
  \varepsilon^{\rm kin}_{\alpha}(\vec{r};t)&=\int\!\!d\vec{p}\,\frac{p^{2}}{2m_{\alpha}}f_{\alpha}(\vec{r},\vec{p};t), \\ \nonumber
  \varepsilon^{\rm int}_{\alpha}(\vec{r};t)&=\sum_{\gamma}\int\!\!d\vec{p}\int\!\!d\vec{p}'\int\!\!d\vec{r}'\,
  \Phi_{\alpha\gamma}(\vec{r},\vec{r}')f_{\alpha\gamma}(\vec{r},\vec{p},\vec{r}',\vec{p}';t)  \nonumber
  +\int\!\!d\vec{p}\int\!\!d\vec{R}_{s}\Phi_{\alpha s}(\vec{r},\vec{R}_{s_{}})
  f_{\alpha s}(\vec{r},\vec{p},\vec{R}_{s};t),
 \end{align*}\\[-5mm]
 which satisfy the corresponding laws of conservation of average nonequilibrium values of the number of ions $n_{\alpha}(\vec{r};t)$,
 full momentum $\vec{p}(\vec{r};t)$\vspace{-2mm}
 \begin{equation*} %\label{e.072}
  \vec{p}(\vec{r};t)=\sum_{\alpha} \vec{p}_{\alpha}(\vec{r};t)
 \end{equation*}\\[-5mm]
 and full of energy
 \begin{equation*} %\label{e.073}
  \varepsilon(\vec{r};t)
  =
  \sum_{\alpha}\big(\varepsilon^{\rm kin}_{\alpha}(\vec{r};t)+\varepsilon^{\rm int}_{\alpha}(\vec{r};t)\big),
 \end{equation*}\\[-3mm]
 underlying the hydrodynamic description of nonequilibrium processes in the system ionic solution -- porous medium.
 The system of equations (\ref{e.03})--(\ref{e.04a}) for nonequilibrium ion distribution functions includes paired quasi-equilibrium coordinate functions ion distribution (positively and negatively charged ions)
 $g_{\alpha \gamma}(\vec{r}, \vec{r}'|n, \beta; t)$
 ($g_{++}$, $g_{+ -}$, $g_{- -}$),
 which are functions of nonequilibrium ion densities $n_{\alpha}(\vec{r};t)$ and inverse temperature $\beta(\vec{r};t)$.
 They describe multiparticle correlations and may be of independent interest,
 as they may be related to the corresponding partial dynamic structural factors of the corresponding quasi-equilibrium state.

 The calculation of paired quasi-equilibrium distribution functions is one of the important problems.
 In particular,
 in the case of simple fluids in the works~\cite{Polewczak1,Polewczak33,Karkheck3},
 when considering the corresponding models of the collision integral for this distribution function we used the generalization of the virial decomposition by the density chosen for the time-dependent density.
 Another way is suggested in the recent work~\cite{TokarchukN},
 in which the paired quasi-equilibrium coordinate distribution function of a simple liquid is calculated from the statistical sum of the corresponding quasi-equilibrium particle distribution in the method of collective variables~\cite{19c,18c,20c,hlush,YukhHlush}.
 In addition,
 it is important to note that the use of the Ornstein--Zernicke equation,
 which depends on the time~\cite{Ramsh1,Eu,Eu1,Bra},
 is promising to calculate $g_{2}(\vec{r}_{1},\vec{r}_{2}|n,\beta;t)$.
 In the case of equilibrium spatially homogeneous and inhomogeneous ionic and ionic-molecular systems,
 the methods for solving the Ornstein--Zernike equations are perfectly developed~\cite{Holovko1,Holovko2,Holovko3,Holovko4}.

 In our case,
 the quasi-equilibrium distribution functions are even $g_{\alpha\gamma}(\vec{r}, \vec{r}'|n,\beta;t)$ can be related to full correlation functions $h_{\alpha\gamma}(\vec{r},\vec{r}'|n,\beta;t)$ ($h_{++}$, $h_{+ -}$, $h_{- -}$) that satisfy inhomogeneous Ornstein--Zernike--type equations that depend on time:\vspace{-2mm}
 \begin{equation} \label{e.074}
  h_{\alpha\gamma} (\vec{r}, \vec{r}';t) =c_{\alpha\gamma} (\vec{r}, \vec{r}';t)
  +\sum_{\nu}
  \int d\vec{r}''c_{\alpha\nu}(\vec{r}, \vec{r}'';t) n_{\nu}(\vec{r}'';t)
  h_{\nu\gamma}(\vec{r}'', \vec{r}';t),
 \end{equation}\\[-5mm]
 where $c_{\alpha\gamma}(\vec{r},\vec{r}';t)$ are direct correlation functions of solution ion distribution.
 The system of equations~(\ref{e.074}) includes time-dependent ion densities $n_{\nu}(\vec{r}''; t)$ ($n_{+}$, $n_{-}$),
 equations for which it is necessary to find from the corresponding kinetic equation for $f_{\alpha}(\vec{p},\vec{r};t)$ after integration by impulses $d\vec{p}$.

 It is important to note that the boundary condition in the equation~(\ref{e.04}) takes into account spatially inhomogeneous correlations between solution particles.
 In contrast,
 the boundary condition in the equation~(\ref{e.05}) does not take into account spatially inhomogeneous correlations between solution particles and a porous matrix,
 which corresponds to the principle of complete weakening of Bogolyubov correlations and,
 in this case,
 is a consequence of the independence of the ionic solution from the porous matrix at the initial time.

 \subsection{Kinetic equations with the initial condition of interacting subsystems: \\ ionic solution and porous matrix}

 In the case when at the initial moment of time,
 the ionic solution and the porous matrix are considered by interacting subsystems,
 the first equations of the BBGKI chain have the following form:\vspace{-2mm}
 \begin{equation*} %\label{e.0303}
  \left(\frac{\partial}{\partial t} +iL_{\alpha}(1)\right)f_{\alpha}(x_{1};t)
  +\sum_{\gamma}\int\!\!dx_{2}\,iL_{\alpha\gamma}(1,2) f_{\alpha\gamma}(x_{1},x_{2};t)
  +\int\!\!d\vec{R}_{s}\,iL_{\alpha s}(1,s) f_{\alpha s}(x_{1},\vec{R}_{s};t)=0,
 \end{equation*}\\[-13mm]
 \begin{multline*} %\label{e.0404}
  \left(\frac{\partial }{\partial t}+iL_{\alpha}(1)+iL_{\gamma}(2)+iL_{\alpha\gamma}(1,2)\right)f_{\alpha\gamma}(x_{1},x_{2};t) \\
   + \sum_{\xi}\int\!\!dx_{3} \big(iL_{\alpha\xi}(1,3) +iL_{\gamma\xi}(2,3)\big)
   f_{\alpha\gamma\xi}(x_{1},x_{2},x_{3};t)
  + \int \!\!d\vec{R}_{s}\big(iL_{\alpha s}(1,s) +iL_{\gamma s}(2,s)\big)
  f_{\alpha\gamma s}(x_{1},x_{2},\vec{R}_{s};t) \\
  =- \varepsilon \big(f_{\alpha\gamma}(x_{1},x_{2};t)
   - g_{\alpha\gamma}(\vec{r}_{1},\vec{r}_{2}|n,\beta;t) f_{\alpha}(x_{1};t)f_{\gamma}(x_{2};t)\big),
 \end{multline*}\\[-16mm]
 \begin{multline*} %\label{e.04a4}
  \left(\frac{\partial }{\partial t} +iL_{\alpha}(1)+iL_{\alpha s}(1,s)\right)
  f_{\alpha s}(x_{1},\vec{R}_{s};t) \\
  +  \sum_{\gamma}\int dx_{3}\Big(iL_{\alpha\gamma}(1,3)
  +iL_{s\gamma}(s,3)\Big) f_{\alpha s \gamma}(x_{1},\vec{R}_{s},x_{3};t)
  +\int\!\!d\vec{R}_{s'}iL_{\alpha s'}(1,s')
  f_{\alpha ss'}(x_{1},\vec{R}_{s},\vec{R}_{s'};t) \\
  =-\varepsilon \big( f_{\alpha s}(x_{1},\vec{R}_{s};t)- g_{\alpha s}(\vec{r}_{1},
  \vec{R}_{s}|n,\beta;t)f_{\alpha}(x_{1};t)n(\vec{R}_{s};t)\big),
 \end{multline*}\\[-7mm]
 where $g_{\alpha\gamma}(\vec{r}_{1},\vec{r}_{2}|n,\beta;t)$,
 $g_{\alpha s}(\vec{r}_{1},\vec{R}_{s}|n,\beta;t)$
 are quasi-equilibrium pair ion distribution functions of $\alpha,\gamma$ varieties solution subsystems and solution -- porous matrix,
 $n(\vec{r}_{s};t)$ is a nonequilibrium density
 (unary function distribution)
 of the porous subsystem particles,
 $f_{\alpha}(x_{1};t)$,
 $f_{\alpha\gamma}(x_{1},x_{2};t)$,
 $f_{\alpha\gamma\xi}(x_{1},x_{2},x_{3};t)$
 are one-, two- and three-ion nonequilibrium distribution functions,
 $f_{\alpha\gamma s}(x_{1},x_{2},\vec{R}_{s};t)$,
 $f_{\alpha s}(x_{1},\vec{R}_{s};t)$,
 $f_{\alpha ss'}(x_{1},\vec{R}_{s},\vec{R}_{s'};t)$ are nonequilibrium distribution functions ions and particles of the porous medium.
 Paired quasi-equilibrium distribution functions
 $g_{\alpha\gamma}(\vec{r},\vec{r}'|n,\beta;t)$,
 $g_{\alpha s}(\vec{r},\vec{R}_{s}|n,\beta;t)$
 in the chain of equations BBGKI as in the previous case can be associated with the corresponding complete correlation functions $h_{\alpha\gamma}(\vec{r},\vec{r}'|n,\beta;t)$
 ($h_{++}$, $h_{- +}$, $h_{- - }$),
 $h_{\alpha s}(\vec{r},\vec{R}_{s}|n,\beta;t)$
 ($h_{+ s}$, $h_{- s}$),
 which satisfy inhomogeneous equations Ornstein--Zernike,
 which depend on time:\vspace{-3mm}
\begin{multline*} %\label{e.0740}
  h_{\alpha\gamma} (\vec{r}, \vec{r}';t) =c_{\alpha\gamma} (\vec{r}, \vec{r}';t)
  +\sum_{\xi}
  \int d\vec{r}''c_{\alpha\xi}(\vec{r}, \vec{r}'';t) n_{\xi}(\vec{r}'';t)
  h_{\xi\gamma}(\vec{r}'', \vec{r}';t)\\
  +  \int d\vec{R}_{s}\,c_{\alpha s}(\vec{r}, \vec{R}_{s};t) n_{s}(\vec{R}_{s};t)h_{s \gamma}(\vec{R}_{s}, \vec{r}';t),
 \end{multline*}\\[-13mm]
 \begin{multline*} %\label{e.075}
  h_{\alpha s} (\vec{r}, \vec{R}_{s};t) =c_{\alpha s} (\vec{r}, \vec{R}_{s};t)
  +\sum_{\xi}  \int d\vec{r}''c_{\alpha\xi}(\vec{r}, \vec{r}'';t) n_{\xi}(\vec{r}'';t)
  h_{\xi s}(\vec{r}'', \vec{R}_{s};t)\\
  +  \int d\vec{R}_{s'}\,c_{\alpha s}(\vec{r}, \vec{R}_{s};t) n_{s'}(\vec{R}_{s'};t)
  h_{s' s}(\vec{R}_{s'}, \vec{R}_{s}),
 \end{multline*}\\[-13mm]
 \begin{multline*} %\label{e.0075}
  h_{s \alpha} (\vec{R}_{s}, \vec{r};t) =c_{s \alpha} (\vec{R}_{s}, \vec{r};t)
  +\sum_{\xi}
  \int d\vec{r}''c_{s \xi}(\vec{R}_{s}, \vec{r}'';t) n_{\xi}(\vec{r}'';t)
  h_{\xi \alpha}(\vec{r}'', \vec{r};t)\\
  +  \int d\vec{R}_{s'}\,c_{s s'}(\vec{R}_{s}, \vec{R}_{s'}) n_{s'}(\vec{R}_{s'};t)
  h_{s' \alpha}(\vec{R}_{s'}, \vec{r};t),
 \end{multline*}
 \begin{equation*} %\label{e.0076}
  h_{s s} (\vec{R}_{s}, \vec{R}_{s'};t) =c_{s s} (\vec{R}_{s}, \vec{R}_{s'};t)
  +\int d\vec{R}_{s''}c_{s s}(\vec{R}_{s}, \vec{R}_{s''};t) n_{s}(\vec{R}_{s''};t)
   h_{ss}(\vec{R}_{s''}, \vec{R}_{s'};t),
 \end{equation*}
 where $c_{\alpha\gamma}(\vec{r},\vec{r}';t)$,
 $c_{\alpha s}(\vec{r},\vec{R}_{s};t)$,
 $c_{s s}(\vec{R}_{s},\vec{R}_{s'};t)$
 are direct correlation functions of ions and particles of the porous medium.
 As we can see,
 the structure is a connected system of equations for finding the correlation functions
 $h_{\alpha\gamma}(\vec{r}, \vec{r}'|n,\beta;t)$,
 $h_{\alpha s}(\vec{r}, \vec{R}_{s}|n,\beta;t)$.
 It includes the nonequilibrium density $n_{s}(\vec {R}_{s''}; t)$,
 $h_{s s}(\vec{R}_{s},\vec{R}_{s'};t)$,
 $c_{s s}(\vec{R}_{s},\vec{R}_{s'};t)$ particles of a porous matrix.
 This raises the problem of developing methods for solving time-dependent inhomogeneous Ornstein--Zernike equations.

 In the next section,
 we consider the approximation of pair collisions between particles for the case when the subsystems ionic solution and the porous matrix at the initial time are considered interacting.

 \section{Approaching pairwise collisions}

 In the approximate pair collisions between particles,
 when three-particle distribution functions are not taken into account,
 for nonequilibrium two-particle distribution functions,
 we obtain~\cite{l1,zmot} the following equations:\vspace{-3mm}
 \begin{multline} \label{e.07}
  \left(\frac{\partial }{\partial t} +iL_{\alpha}(1)+iL_{\gamma}(2)+iL_{\alpha\gamma}(1,2)\right)
   f_{\alpha\gamma}(x_{1},x_{2};t)     \\
   = -\varepsilon \big(f_{\alpha\gamma}(x_{1},x_{2};t)
     -g_{\alpha\gamma}(\vec{r}_{1}, \vec{r}_{2}|n,\beta;t)f_{\alpha}(x_{1};t)f_{\gamma}(x_{2};t)\big),
 \end{multline}\\[-15mm]
 \begin{multline} \label{e.08}
  \left(\frac{\partial }{\partial t} +iL_{\alpha}(1)+iL_{\alpha s}(1,s)\right)f_{\alpha s}(x_{1},\vec{R}_{s};t)   \\
   = -\varepsilon \big(f_{\alpha s}(x_{1},\vec{R}_{s};t)
     -g_{\alpha s}(\vec{r}_{1}, \vec{R}_{s}|n,\beta;t)f_{\alpha}(x_{1};t)n(\vec{R}_{s};t)\big).
 \end{multline}\\[-7mm]
 Solutions of these equations (\ref{e.07}), (\ref{e.08}) can be given as:\vspace{-2mm}
 \begin{align*} %\label{e.09}
  f_{\alpha\gamma}(x_{1},x_{2};t)&=\varepsilon \int_{-\infty}^{0} \!\!d\tau\,
  e^{\big(\varepsilon +iL^{(2)}_{\alpha\gamma}(1,2)\big)\tau}
  g_{\alpha\gamma}(\vec{r}_{1},\vec{r}_{2}|n,\beta;t+\tau)
   f_{\alpha}(x_{1};t+\tau)f_{\gamma}(x_{2};t+\tau),  \\ % \label{e.010}
  f_{\alpha  s}(x_{1},\vec{R}_{s};t)&=\varepsilon \int_{-\infty}^{0}\!\!
  d\tau \,e^{\big(\varepsilon +iL^{(2)}_{\alpha s}(1,s)\big)\tau} g_{\alpha s}(\vec{r}_{1},
  \vec{R}_{s}|n,\beta;t+\tau) f_{\alpha}(x_{1};t+\tau)n(\vec{R}_{s};t+\tau),
 \end{align*}\\[-5mm]
 where\vspace{-3mm}
 \begin{align*}
  iL^{(2)}_{\alpha\gamma}(1,2)&=iL_{\alpha}(1)+iL_{\gamma}(2)+iL_{\alpha\gamma}(1,2),\\
  iL^{(2)}_{\alpha s}(1,s)    &= iL_{\alpha}(1)+iL_{\alpha s}(1,s).
 \end{align*}\\[-7mm]
 Substituting these solutions into the equation~(\ref{e.03}),
 we obtain a non-Markov kinetic equation for nonequilibrium one-particle ion distribution function in the system ionic solution -- porous matrix:\vspace{-3mm}
 \begin{multline} \label{e.011}
  \left(\frac{\partial }{\partial t} +iL_{\alpha}(1)\right)f_{\alpha}(x_{1};t)=
 -\sum_{\gamma}\int dx_{2}\,iL_{\alpha\gamma}(1,2) \\
 \times  \varepsilon \int_{-\infty}^{0}\!\! d\tau \,
 e^{\big(\varepsilon +iL^{(2)}_{\alpha\gamma}(1,2)\big)\tau}
  g_{\alpha\gamma}(\vec{r}_{1}, \vec{r}_{2}|n,\beta;t+\tau)
   \, f_{\alpha}(x_{1};t+\tau)f_{\gamma}(x_{2};t+\tau)  \\
  -  \int d\vec{R}_{s}\,iL(1,s)\,\varepsilon \int_{-\infty}^{0} \!\!
  d\tau \,e^{\big(\varepsilon +iL^{(2)}_{\alpha s}(1,s)\big)\tau} g_{\alpha s}(\vec{r}_{1},
 \vec{R}_{s}|n,\beta;t+\tau)
  \, f_{\alpha}(x_{1};t+\tau)n(\vec{R}_{s};t+\tau).
 \end{multline}\\[-9mm]

 The obtained kinetic equation for the nonequilibrium one-particle ion distribution function also takes into account the spatial heterogeneity of the system.
 If this equation formally put $g_{\alpha s}(\vec{r}_{1},\vec{R}_{s}|n,\beta;t+\tau)=1$,
 then we obtain the kinetic equation for the nonequilibrium one-particle distribution function ions,
 which corresponds to the initial condition of non-interacting subsystems of ionic solution and porous matrix.
 When obtaining the kinetic equation~(\ref{e.011}),
 we did not specify the paired potentials of interaction between ions and particles of the porous matrix.

 In the next section,
 we consider the model of charged solid spheres to describe the ionic solution in the presence of a porous matrix.

 \section{Model of charged solid spheres.
          Kinetic equation of the revised Enskog--Vlasov--Landau theory for the ionic solution-porous matrix system}

 Consider the model of charged solid spheres for an ionic subsystem when the interaction potential can be submitted as an amount~\cite{l1,zmot,kobr2}:\vspace{-2mm}
 \[
  \Phi_{\alpha\gamma}(\vec{r},\vec{r}')=
  \Phi^{sh}_{\alpha\gamma}(\vec{r},\vec{r}')
  +\Phi^{l}_{\alpha\gamma}(\vec{r},\vec{r}'),
 \]\\[-7mm]
 where $\Phi^{\rm sh}_{\alpha\gamma}(\vec{r},\vec{r}')$ is the short-range interaction potential between ions,
 which will be modeled by the sum of $\Phi^{\text{sh-rep}}_{\alpha\gamma}(\vec{r},\vec{r}')$ is the potential of solid spheres and $\Phi^{\text{sh-att}}_{\alpha\gamma}(\vec{r},\vec{r}')$ is short-range attraction potential,
 which describes possible associative connections between ions;
 $\Phi^{l}_{\alpha\gamma}(\vec{r},\vec{r}')$ is a long-range potential for interaction between ions,
 in particular the potential of the Yukawa type.
 In addition,
 the interaction of ions and particles of the porous medium will be described by short-term potential of solid spheres
 $\Phi^{\text{sh-rep}}_{\alpha s}(\vec{r},\vec{R}_{s})$ and some attractive potential
 $\Phi^{\text{sh-att}}_{\alpha s}(\vec{r},\vec{R}_{s})$ with effective range $r_{eff}$.
 Based on the works~\cite{l1,zmot} in the case of the model of solid spheres for the liquid subsystem from~(\ref{e.011}) we get:\vspace{-2mm}
 \begin{multline*} %\label{e.1} %\label{e.a0}
  \left(\frac{\partial }{\partial t} +iL_{\alpha}(1)\right)
  f_{\alpha}(x_{1};t)
  =-\sum_{\gamma}\int_{0}^{\sigma_{\gamma}}\!\!\! d\vec{r}_{2}\int d\vec{p}_{2} \,
   iL_{\alpha\gamma}^{\text{sh-rep}}(12) \\[-1mm]
   \times \varepsilon \int_{-\infty}^{0}\!\!\!\!\!\!d\tau \,e^{\big(\varepsilon
   +iL_{\alpha\gamma}^{0}(12)+iL_{\alpha\gamma}^{\text{sh-rep}}(12)\big)\tau}
   g_{\alpha\gamma}(\vec{r}_{1}, \vec{r}_{2}|n,\beta;t+\tau)
   \, f_{\alpha}(x_{1};t+\tau)f_{\gamma}(x_{2};t+\tau)  \\[-1mm]
  -\sum_{\gamma}\int_{\sigma_{\gamma}}^{r_{\rm eff}} \!\!\!\!\!\!d\vec{r}_{2}\int \!\!d\vec{p}_{2}\,
  iL_{\alpha\gamma}^{\text{sh-att}}(12)\,\varepsilon
  \int_{-\infty}^{0}\!\!\!\!\!\!d\tau\,e^{\big(\varepsilon +iL_{\alpha\gamma}^{0}(12)+iL_{\alpha\gamma}^{\text{sh-att}}(12)\big)\tau} \\
  \times g_{\alpha\gamma}(\vec{r}_{1},\vec{r}_{2}|n,\beta;t+\tau)f_{\alpha}(x_{1};t+\tau)f_{\gamma}(x_{2};t+\tau) \\
  -\sum_{\gamma}\int_{r_{\rm eff}}^{\infty}\!\!\!\!\!d\vec{r}_{2}\int \!\!d\vec{p}_{2}\,
  iL_{\alpha\gamma}^{l}(12)\,\varepsilon \!\!\int_{-\infty}^{0}\!\!\!\!\!\!d\tau\,
  e^{\big(\varepsilon +iL_{\alpha\gamma}^{0}(12)+iL_{\alpha\gamma}^{l}(12)\big)\tau}
  g_{\alpha\gamma}(\vec{r}_{1},\vec{r}_{2}|n,\beta;t+\tau)f_{\alpha}(x_{1};t+\tau)f_{\gamma}(x_{2};t+\tau) \\[-1mm]
  -\int_{0}^{\sigma_{s}}\!\!\!\!d\vec{R}_{s} \,iL_{\alpha s}^{\text{sh-rep}}(1s)\,
  \varepsilon \int_{-\infty}^{0}\!\!\!\!\!\!d\tau\, e^{\big(\varepsilon +iL_{\alpha}(1)+iL_{\alpha s}^{\text{sh-rep}}(1s)\big)\tau}
  g_{\alpha s}(\vec{r}_{1}, \vec{R}_{s}|n,\beta;t+\tau)f_{\alpha}(x_{1};t+\tau)n_{s} (\vec{R}_{s};t+\tau)  \\[-1mm]
  -\int_{\sigma_{s}}^{r_{\rm eff}}\!\!\!\! d\vec{R}_{s} \,iL_{\alpha s}^{\text{sh-att}}(1s)\,
  \varepsilon \int_{-\infty}^{0}\!\!\!\!\!\!d\tau \,e^{\big(\varepsilon +iL_{\alpha}(1)+iL_{\alpha s}^{\text{sh-att}}(1s)\big)\tau}
  g_{\alpha s}(\vec{r}_{1}, \vec{R}_{s}|n,\beta;t+\tau)f_{\alpha}(x_{1};t+\tau)n_{s} (\vec{R}_{s};t+\tau)
 \end{multline*}\\[-4mm]
 is a kinetic equation for the nonequilibrium one-particle ion distribution function,
 taking into account areas of short-range (solid-sphere) and long-range potentials.
 Given that in the field of action of the potential of solid spheres,
 the interaction time $\tau\rightarrow+0$ and detailed calculations~\cite{l1,zmot,kobr2},
 the equation can be given as:\vspace{-3mm}
 \begin{multline*} %\label{e.a1}
  \left(\frac{\partial }{\partial t} +iL_{\alpha}(1)\right)
  f_{\alpha}(x_{1};t)
  =
  - \sum_{\gamma}\int dx_{2} \,\hat{T}_{\alpha\gamma}(12)\,g_{\alpha\gamma}(\vec{r}_{1},
    \vec{r}_{2}|n,\beta;t) \,  f_{\alpha}(x_{1};t)f_{\gamma}(x_{2};t)\\[-1mm]
-\int d\vec{R}_{s}\, \hat{T}_{\alpha s}(1s)\,g_{\alpha s}(\vec{r}_{1},
\vec{R}_{s}|n,\beta;t)f_{\alpha}(x_{1};t)n_{s} (\vec{R}_{s};t) \\[-1mm]
  -\sum_{\gamma}\int_{\sigma_{\gamma}}^{r_{\rm eff}} \!\!\!\!\!\!d\vec{r}_{2}
  \int\!\!d\vec{p}_{2} \,iL_{\alpha\gamma}^{\text{sh-att}}(12)\,\varepsilon \!
  \int_{-\infty}^{0}\!\!\!\!\!\!d\tau \,e^{(\varepsilon +iL_{\alpha\gamma}^{0}(12)+iL_{\alpha\gamma}^{\text{sh-att}}(12))\tau}\\[-1mm]
\times \,\, g_{\alpha\gamma}(\vec{r}_{1},
  \vec{r}_{2}|n,\beta;t+\tau)f_{\alpha}(x_{1};t+\tau)f_{\gamma}(x_{2};t+\tau) \\[-1mm]
  -\sum_{\gamma}\int_{r_{\rm eff}}^{\infty}\!\!\!\!\!\!d\vec{r}_{2}
  \int\!\!d\vec{p}_{2}\,iL_{\alpha\gamma}^{l}(12)\,\varepsilon \!
  \int_{-\infty}^{0}\!\!\!\!\!\!d\tau
  e^{(\varepsilon +iL_{\alpha\gamma}^{0}(12)+iL_{\alpha\gamma}^{l}(12))\tau}
  g_{\alpha\gamma}(\vec{r}_{1},
\vec{r}_{2}|n,\beta;t+\tau)f_{\alpha}(x_{1};t+\tau)f_{\gamma}(x_{2};t+\tau)\\[-1mm]
 -\int_{\sigma_{s}}^{r_{\rm eff}}\!\!\!\!d\vec{R}_{s} \,iL_{\alpha s}^{\text{sh-att}}(1s)\,
  \varepsilon
  \int_{-\infty}^{0}\!\!\!\!\!\!d\tau \,e^{(\varepsilon +iL_{\alpha}(1)+iL_{\alpha s}^{\text{sh-att}}(1s))\tau}
  g_{\alpha s}(\vec{r}_{1},\vec{R}_{s}|n,\beta;t+\tau)f_{\alpha}(x_{1};t+\tau)
  n_{s}(\vec{R}_{s};t+\tau),
 \end{multline*}\\[-4mm]
 where $\hat{T}_{\alpha\gamma}(12)$ is the Enskog collision operator for charged solid spheres (ions)~\cite{l1}, $\hat{T}_{\alpha s}(1s)$ is the operator Enskog collision for charged solid spheres and solid spheres,
 describing the porous medium.
 Next,
 if in the long-range part of the collision integral to perform integration by parts,
 we obtain the following kinetic equation:\vspace{-3mm}
 \begin{multline*} %\label{e.aa1}
  \left(\frac{\partial }{\partial t} +iL_{\alpha}(1)\right)f_{\alpha}(x_{1};t)
  =-\sum_{\gamma}\int dx_{2}\,
  \hat{T}_{\alpha\gamma}(12)\,g_{\alpha\gamma}(\vec{r}_{1},\vec{r}_{2}|n,\beta;t)\,f_{\alpha}(x_{1};t)\,f_{\gamma}(x_{2};t)\\
  -\int d\vec{R}_{s} \,\hat{T}_{\alpha s}(1s)\,g_{\alpha s}(\vec{r}_{1},\vec{R}_{s}|n,\beta;t)\,f_{\alpha}(x_{1};t)\,
  n_{s} (\vec{R}_{s};t)   \\
  -\sum_{\gamma}\int_{\sigma_{\gamma}}^{r_{\rm eff}} d\vec{r}_{2}
  \int d\vec{p}_{2}\,iL_{\alpha\gamma}^{\text{sh-att}}(12)\,g_{\alpha\gamma}(\vec{r}_{1},\vec{r}_{2}|n,\beta;t)\,
  f_{\alpha}(x_{1};t)\,f_{\gamma}(x_{2};t)  \\
  +\sum_{\gamma}\int_{\sigma_{\gamma}}^{r_{\rm eff}} d\vec{r}_{2}
  \int d\vec{p}_{2} \,iL_{\alpha\gamma}^{\text{sh-att}}(12)\int_{-\infty}^{0}d\tau\,
  e^{(\varepsilon +iL_{\alpha\gamma}^{0}(12)+iL_{\alpha\gamma}^{\text{sh-att}}(12))\tau} \\
  \times\left(\frac{\partial }{\partial \tau}+iL_{\alpha\gamma}^{0}(12)+iL_{\alpha\gamma}^{\text{sh-att}}(12)\right)
  g_{\alpha\gamma}(\vec{r}_{1},\vec{r}_{2}|n,\beta;t+\tau)\,
  f_{\alpha}(x_{1};t+\tau)\,
  f_{\gamma}(x_{2};t+\tau)  \\
  -\sum_{\gamma}\int_{r_{\rm eff}}^{\infty} d\vec{r}_{2}
  \int d\vec{p}_{2} \,iL_{\alpha\gamma}^{l}(12)\,g_{\alpha\gamma}(\vec{r}_{1},\vec{r}_{2}|n,\beta;t)\,
  f_{\alpha}(x_{1};t)\,
  f_{\gamma}(x_{2};t)  \\
  +\sum_{\gamma}\int_{r_{\rm eff}}^{\infty} d\vec{r}_{2}
  \int d\vec{p}_{2} \,iL_{\alpha\gamma}^{l}(12)\int_{-\infty}^{0}d\tau\,
  e^{(\varepsilon +iL_{\alpha\gamma}^{0}(12)+iL_{\alpha\gamma}^{l}(12))\tau} \\
  \times\left(\frac{\partial }{\partial \tau}+iL_{\alpha\gamma}^{0}(12)+iL_{\alpha\gamma}^{l}(12)\right)
  g_{\alpha\gamma}(\vec{r}_{1},\vec{r}_{2}|n,\beta;t+\tau)\,
  f_{\alpha}(x_{1};t+\tau)\,
  f_{\gamma}(x_{2};t+\tau)  \\
  -\int_{\sigma_{s}}^{r_{\rm eff}} d\vec{R}_{s} \,iL_{\alpha s}^{\text{sh-att}}(1s)
  g_{\alpha s}(\vec{r}_{1},\vec{R}_{s}|n,\beta ;t)\,
  f_{\alpha}(x_{1};t)\,
  n_{s} (\vec{R}_{s};t)  \\
  +\int_{\sigma_{s}}^{r_{\rm eff}} d\vec{R}_{s} \,iL_{\alpha s}^{\text{sh-att}}(1s)
  \int_{-\infty}^{0}d\tau \,e^{(\varepsilon +iL_{\alpha}(1)+iL_{\alpha s}^{\text{sh-att}}(1s))\tau}  \\
  \times \left(\frac{\partial }{\partial \tau}+iL_{\alpha}(1)+iL_{\alpha s}^{\text{sh-att}}(1s)\right)
  g_{\alpha s}(\vec{r}_{1},\vec{R}_{s}|n,\beta;t+\tau)\,
  f_{\alpha}(x_{1};t+\tau)\,
  n_{s} (\vec{R}_{s};t+\tau),
 \end{multline*}\\[-5mm]
 where the third,
 fifth,
 and seventh terms in the right part are generalized integrals of the Vlasov collision -- generalized middle fields,
 and the fourth,
 sixth,
 and eighth terms are generalized Landau--type collision integrals between ions and ions and porous matrix particles,
 taking into account memory effects.

 Revealing the action of the operator Enskog in the right part,
 in a spatially inhomogeneous case,
 (up to linear values by gradients) and without taking into account the effects of memory,
 we obtain:\vspace{-2mm}
 \begin{equation} \label{e.2}
  \left(\frac{\partial }{\partial t} +iL_{\alpha}(1)\right)f_{\alpha}(x_{1};t)
  =
  I^{(0)}_{\alpha E}(x_{1};t) + I^{(1)}_{\alpha E}(x_{1};t) +
  I^{(1)}_{\alpha MF}(x_{1};t) +I^{(1)}_{\alpha L}(x_{1};t),
 \end{equation}\\[-5mm]
 where the terms on the right are the integrals of collisions,
 due to the contribution of an certain type of interparticle interaction.
 The first and second of them are Enskog--type collision integrals of the RET theory~\cite{l1}:\vspace{-4mm}
 \begin{multline*} %\label{e.3}
  I^{(0)}_{\alpha E}(x_{1};t)=\sum_{\gamma}\int d\vec{v}_{2}\int d\varepsilon
  \int b \,db\, g(12)g_{\alpha\gamma}(\sigma_{\alpha\gamma}|n, \beta ;t) \\[-2mm]
  \times
  \big(f_{\alpha}(\vec{r}_{1},\vec{v}'_{1};t)f_{\gamma}(\vec{r}_{2},\vec{v}'_{2};t)
  -f_{1}(\vec{r}_{\alpha},\vec{v}_{1};t)f_{\gamma}(\vec{r}_{2},\vec{v}_{2};t)\big),
 \end{multline*}\\[-16mm]
 \begin{multline*} %\label{e.4}
  I^{(1)}_{\alpha E}(x_{1};t)=\sum_{\gamma}\sigma_{\alpha\gamma}^{3}\int d\hat{\vec{r}}_{12}
  \int d\vec{v}_{2}\,
  \Theta \big(\hat{\vec{r}}_{12}\cdot \vec{g}(12)\big)
  \big(\hat{\vec{r}}_{12}\cdot \vec{g}(12)\big)      \\[-1mm]
  \times \Big(g_{\alpha\gamma}(\vec{r}_{12}|n;t)\vec{r}_{12}
\cdot[f_{\alpha}(\vec{r}_{1},\vec{v'}_{1};t)\vec{\nabla}_{2}
 f_{\gamma}(\vec{r}_{2},\vec{v'}_{2};t) -f_{\alpha}(\vec{r}_{1},\vec{v}_{1};t)\vec{\nabla}_{2}
f_{\gamma}(\vec{r}_{2},\vec{v}_{2};t)]    \\[-1mm]
+\frac{1}{2}\big(\hat{\vec{r}}_{12}\cdot \vec{\nabla}_{2}
g_{\alpha\gamma}(\vec{r}_{12} |n, \beta ;t)\big)
\big[f_{\alpha}(\vec{r}_{1},\vec{v'}_{1};t)f_{\gamma}(\vec{r}_{2},\vec{v'}_{2};t)-
f_{\alpha}(\vec{r}_{1},\vec{v}_{1};t)f_{\gamma}(\vec{r}_{2},\vec{v}_{2};t)\big]\Big),
 \end{multline*}\\[-7mm]
 where $b$ is the aiming parameter,
 $g_{\alpha\gamma}(\sigma_{\alpha\gamma}|n;t)$ is the contact value of the paired quasi-equilibrium distribution function,
 $\hat{\vec{r}}_{12}=\frac{\vec{r}_{12}}{|\vec{r}_{12}|}$ is a single vector,
 $\vec{v'}_{1}=\vec{v}_{1}+\hat{\vec{r}}_{12}(\hat{\vec{r}}_{12}\cdot \vec{g}(12))$,
 $\vec{v'}_{2}=\vec{v}_{2}-\hat{\vec{r}}_{12}(\hat{\vec{r}}_{12}\cdot \vec{g}(12))$
 is the value of particle velocities $1$, $2$ after the collision,
 while $\vec{v}_{1}$,
 $\vec{v}_{2}$ is the value of their velocities before the collision,
 where $\vec{g}(12)=\vec{v}_{2}-\vec{v}_{1}$ is the relative velocity.

 The next term is the contribution of the mean-field theory KMFT~\cite{l1}:\vspace{-3mm}
 \begin{multline*} %\label{e.3a}
  I^{(1)}_{\alpha MF}(x_{1};t)=\frac{1}{m_{\alpha}} \sum_{\gamma} \int_{\sigma_{\gamma}}^{r_{\rm eff}} d\vec{r}_{2}\,
  \frac{\partial }{\partial \vec{r}_{1}} \Phi^{\text{sh-att}}_{\alpha\gamma}(\vec{r}_{12})
  \cdot\frac{\partial }{\partial \vec{v}_{1}}
  g_{\alpha\gamma}(\vec{r}_{12} |n,\beta;t) f_{\alpha}(\vec{r}_{1},\vec{v}_{1};t)n_{\gamma}(\vec{r}_{2};t) \\
  +\frac{1}{m_{\alpha}} \sum_{\gamma} \int_{r_{\rm eff}}^{\infty} d\vec{r}_{2}\,
    \frac{\partial }{\partial \vec{r}_{1}} \Phi^{l}_{\alpha\gamma}(\vec{r}_{12})
    \cdot
    \frac{\partial }{\partial \vec{v}_{1}} g_{\alpha\gamma}(\vec{r}_{12} |n,\beta;t)
    f_{\alpha}(\vec{r}_{1},\vec{v}_{1};t)n_{\gamma}(\vec{r}_{2};t)\\
  +\frac{1}{m_{\alpha}} \sum_{\gamma} \int_{\sigma_{\gamma}}^{r_{\rm eff}} d\vec{R}_{2}\,
   \frac{\partial }{\partial \vec{r}_{1}} \Phi^{\text{sh-att}}_{\alpha s}(\vec{r}_{1},\vec{R}_{2},)
   \cdot
   \frac{\partial }{\partial \vec{v}_{1}}
   g_{\alpha s}(\vec{r}_{1},\vec{R}_{2} |n,\beta;t) f_{\alpha}(\vec{r}_{1},\vec{v}_{1};t)n_{s}(\vec{R}_{2};t).
 \end{multline*}\\[-5mm]
 The last term is the integral of Landau--type collisions~\cite{l1,zmot}\vspace{-4mm}
 \begin{multline*} %\label{e.30aa}
  I^{(1)}_{\alpha L}(x_{1};t)=
  \sum_{\gamma}\int_{\sigma_{\gamma}}^{r_{\rm eff}} d\vec{r}_{2}
  \int d\vec{p}_{2} \,iL_{\alpha\gamma}^{\text{sh-att}}(12)\int_{-\infty}^{0}d\tau\,
  e^{(\varepsilon +iL_{\alpha\gamma}^{0}(12)+iL_{\alpha\gamma}^{\text{sh-att}}(12))\tau} \\
  \times
  \left(\frac{\partial }{\partial \tau}+iL_{\alpha\gamma}^{\text{sh-att}}(12)\right)
  g_{\alpha\gamma}(\vec{r}_{1},
  \vec{r}_{2}|n,\beta;t+\tau)f_{\alpha}(x_{1};t+\tau)f_{\gamma}(x_{2};t+\tau) \\
  +\sum_{\gamma}\int_{r_{\rm eff}}^{\infty} d\vec{r}_{2}
  \int d\vec{p}_{2} \,iL_{\alpha\gamma}^{l}(12)\int_{-\infty}^{0}d\tau\,
  e^{(\varepsilon +iL_{\alpha\gamma}^{0}(12)+iL_{\alpha\gamma}^{l}(12))\tau} \\
  \times
  \left(\frac{\partial }{\partial \tau}+iL_{\alpha\gamma}^{l}(12)\right)
  g_{\alpha\gamma}(\vec{r}_{1},\vec{r}_{2}|n,\beta;t+\tau)
  f_{\alpha}(x_{1};t+\tau)f_{\gamma}(x_{2};t+\tau) \\
  +\int_{\sigma_{s}}^{r_{\rm eff}} d\vec{R}_{2}
  iL_{\alpha s}^{\text{sh-att}}(12)
  \int_{-\infty}^{0}d\tau\,
  e^{(\varepsilon +iL_{\alpha s}^{0}(12)+iL_{\alpha s}^{\text{sh-att}}(12))\tau} \\
  \times
  \left(\frac{\partial }{\partial \tau}+iL_{\alpha s}^{\text{sh-att}}(12)\right)
  g_{\alpha s}(\vec{r}_{1},
  \vec{R}_{2}|n,\beta;t+\tau)f_{\alpha}(x_{1};t+\tau)n_{s}(\vec{R}_{2};t+\tau).
 \end{multline*}\\[-5mm]
 By solving~\cite{kobr2} kinetic Enskog--Vlasov--Landau equations~(\ref{e.2}) for charged solid spheres can be constructed as equations hydrodynamics and obtained analytical expressions for mutual diffusion coefficients,
 thermodiffusion,
 viscosity,
 and thermal conductivity through particle distribution functions and them the nature of the interaction.

 \vspace{-2mm}
 \section{Conclusions}\vspace{-1mm}

 The kinetic approach is applied to the description of ion transfer processes in the system ionic solution -- a porous medium.
 The nonequilibrium state of the system is described using a modified one chain of equations BBGKI~\cite{l1,tokom,zmot,tokom01,tokom02} for partial nonequilibrium distribution functions of ions and particles of the porous matrix.
 For this purpose,
 we used the approach proposed in~\cite{l1,tokom,zmot,tokom01,tokom02,TokarchukN},
 where the modified chain of BBGKI equations is built taking into account the concept of the consistent description of kinetics and hydrodynamics of nonequilibrium processes of interacting particles in the Zubarev method of nonequilibrium statistical operator.
 A generalized kinetic equation of the revised Enskog--Vlasov--Landau theory for the nonequilibrium ion distribution function in the model of charged solid spheres is obtained,
 taking into account attractive short-range interactions for the ionic solution -- porous medium system.
 Two cases of construction of kinetic equations depending on the initial conditions of the interaction of ionic solution particles and porous matrix are considered. In the case of non-interacting subsystems (at the initial moment of time),
 the modified boundary conditions of the BBGKI chain of equations do not include quasi-equilibrium coordinate
 (pair and higher-order) functions describing correlations between solution ions and porous matrix particles.
 In the case of interacting subsystems (at the initial moment of time),
 modified boundary conditions in the chain of equations BBGKI include vase-equilibrium coordinate (pair and higher-order) functions describing the correlations between solution ions and porous matrix particles.
 For this case,
 a generalized kinetic equation of the revised Enskog--Vlasov--Landau theory for the nonequilibrium ion distribution function in the model of charged solid spheres is obtained,
 taking into account attractive short-range interactions for the ionic solution -- porous medium system.
 To calculate the paired quasi-equilibrium distribution coordinate functions for ions and particles of the porous matrix,
 it is proposed to use inhomogeneous,
 time-dependent Ornstein--Zernike equations for the corresponding complete correlation functions
 $h_{\alpha\gamma}(\vec{r},\vec{r}'|n,\beta; t)$ ($h_{++}$, $h_{- +}$, $h_{--}$),
 $h_{\alpha s}(\vec{r},\vec{R}_{s}|n,\beta; t)$ ($h_{+ s}$, $h_{- s}$).

 In the following works,
 solutions for the Enskog--Vlasov--Landau kinetic equations for nonequilibrium distribution functions will be found by the boundary conditions method and the corresponding equations of hydrodynamics with generalized ion transfer coefficients in the ionic solution-porous matrix system will be obtained.

%\bibliography{MyBasa2}

\end{document}